\documentclass[conference]{IEEEtran}
\usepackage{pifont}
\usepackage{xspace}
\usepackage{colortbl}
\usepackage{xcolor}
\usepackage{array} 
\newcommand{\namex}{Janus\xspace}
\newcommand{\mypara}[1]{\textbf{\textit{#1}}}
\newcommand{\jing}[1]{\textcolor{black}{#1}}
\usepackage{subfig}

\usepackage[ruled,vlined]{algorithm2e}
\usepackage{booktabs}
\usepackage{cite}
\usepackage{amsmath,amssymb,amsfonts}
\usepackage{algorithmic}
\usepackage{graphicx}
\usepackage{textcomp}

\def\BibTeX{{\rm B\kern-.05em{\sc i\kern-.025em b}\kern-.08em
    T\kern-.1667em\lower.7ex\hbox{E}\kern-.125emX}}

\begin{document}

\title{It Takes Two to Tango: Serverless Workflow Serving via Bilaterally Engaged Resource Adaptation
}
%\title{Efficient Serverless Workflow Serving via Bilaterally Engaged Resource Adaptation}

\author{
\IEEEauthorblockN{
Jing Wu\textsuperscript{1},
Lin Wang\textsuperscript{2}, 
Quanfeng Deng\textsuperscript{1}, 
Chen Yu\textsuperscript{1}, 
Dong Zhang\textsuperscript{3}, 
Bingheng Yan\textsuperscript{3}, 
Fangming Liu\textsuperscript{*1,4}}
\IEEEauthorblockA{\textsuperscript{1}\textit{
National Engineering Research Center for Big Data Technology and System,}\\
\textit{
Services Computing Technology and System Lab, Cluster and Grid Computing Lab,}\\
\textit{
Huazhong University of Science and Technology, China} \\
\textsuperscript{2}\textit{Paderborn University, Germany } \\
\textsuperscript{3}\textit{Inspur Data Co., Ltd., China}\\
\textsuperscript{4}\textit{Peng Cheng Laboratory, China}\\
Email: wujinghust@hust.edu.cn, lin.wang@uni-paderborn.de, quanfengdeng@foxmail.com, \\
yuchen@hust.edu.cn, \{zhangdong, yanbh\}@inspur.com, fangminghk@gmail.com}
% \IEEEcompsocitemizethanks{
% \IEEEcompsocthanksitem \textsuperscript{*§} Corresponding authors.
% }
}

% \author{
% \IEEEauthorblockN{1\textsuperscript{st} Given Name Surname}
% \and
% \and
% \and
% \and

% \and

\maketitle

\begin{abstract}
Serverless platforms typically adopt an early-binding approach for function sizing, requiring developers to specify an immutable size for each function within a workflow beforehand.
Accounting for potential runtime variability, developers must size functions for worst-case scenarios to ensure service-level objectives (SLOs), resulting in significant resource inefficiency. 
To address this issue, we propose Janus, a novel resource adaptation framework for serverless platforms. Janus employs a late-binding approach, allowing function sizes to be dynamically adapted based on runtime conditions.
The main challenge lies in the information barrier between the developer and the provider: developers lack access to runtime information, while providers lack domain knowledge about the workflow. 
To bridge this gap, Janus allows developers to provide hints containing rules and options for resource adaptation. 
Providers then follow these hints to dynamically adjust resource allocation at runtime based on real-time function execution information, ensuring compliance with SLOs. 
We implement Janus and conduct extensive experiments with real-world serverless workflows. 
Our results demonstrate that Janus enhances resource efficiency by up to 34.7\% compared to the state-of-the-art.

\end{abstract}

% \begin{IEEEkeywords}
% component, formatting, style, styling, insert.
% \end{IEEEkeywords}

\section{Introduction}
Serverless computing has become a popular approach for implementing various cloud applications including web services~\cite{asplos23-beehive}, data processing~\cite{socc21-llama,arxiv22-pheromone} and more recently machine learning training/inference~\cite{socc21-morphling,atc21-infaas}. Serverless computing allows the developer to offload infrastructure management tasks to the cloud provider and ensures high resource elasticity through horizontal auto-scaling. Applications developed as serverless workflows can be represented by directed acyclic graphs (DAGs) where a node presents a function and an edge represents the data exchange between functions. When triggered by an event (i.e., a request), the functions will be executed according to the data flow specified by the DAG. 
Moreover, horizontal auto-scaling takes care of the number of function instances based on the real-time request intensity. Yet, the size (e.g., CPU cores and memory size) of each function instance is typically decided with an early-binding approach---the developer sets it according to the service-level objective (SLO), e.g., meeting the end-to-end latency target at the 99th percentile (P99) in the DAG~\cite{osdi22-orion,mac22-wisefuse}.

% Thanks to its advantage of resource auto-scaling,  serverless has rendered it into a promising solution for machine learning inference/training tasks~\cite{socc21-morphling,atc21-infaas}, web services~\cite{asplos23-beehive,asplos23-beehive}, data processing~\cite{socc21-llama,arxiv22-pheromone}, etc.
% These applications are deployed as serverless workflows, with the form of  a directed acyclic graph (DAG) where nodes represent functions and edges express functions' data flow.
% Currently, serverless platforms adopt an early binding serving pattern that requires developers to specify a size (e.g., CPU cores limits or memory limits) for each function, which should not be adjusted throughout its life-cycle.
% Once deployment, developers lose their control.
% Then, providers take over the workflows, and scale out/in (horizontally) function instances, considering real-time workloads (e.g., concurrency~\cite{lambda-concurrency} and CPU/memory usage~\cite{tencent-scalability}), without the privilege to vertically adjust function sizes.

The early-binding approach shoots for the worst case for SLO guarantee and hence leads to considerable resource over-provisioning. Empirically, we observe that the worst-case execution time can be orders of magnitude larger than that of the best case. For example, the gap between the 95th percentile and the 25th percentile of the workflow execution time of Microsoft Durable Functions can be as high as 80 times on average~\cite{osdi22-orion}. Such variability can be attributed to various runtime dynamics including varying input working set size~\cite{socc23-parrotfish,socc22-cypress,eurosys21-ofc,trans-xwj-1,infocom22-o2a,xbh-2,cvpr24-socialcircle,osdi22-orion,mac22-wisefuse} and performance interference~\cite{asplos23-aquatope,osdi20-firm,isca22-lukewarm,socc21-servermore,hpdc23-propack,xwj-2,wosc20-serverless-not-server-less,ic2e22-cpu-tams}. When the size of the function is decided based on the worst case, the resource utilization will be low for most requests. For example, production serverless traces from Huawei Cloud reveal that half of the deployed functions have CPU and memory usage at merely $10\%$ and $19.5\%$, respectively~\cite{socc23-huawei-cloud}. 

% A variety of dynamics at runtime result in remarkable variance of execution performance.
% These dynamics include  varying input working set~\cite{socc23-parrotfish,socc22-cypress,eurosys21-ofc,osdi22-orion,mac22-wisefuse},  performance interference~\cite{asplos23-aquatope,osdi20-firm,isca22-lukewarm,socc21-servermore,hpdc23-propack,wosc20-serverless-not-server-less,ic2e22-cpu-tams}, and unexpected software/hardware failures~\cite{socc22-owl,sosp21-harvest-vm-for-serverless}.
% Empirical study demonstrates that within Microsoft Azure Durable~\cite{azure-durable-function} workflows' execution latency suffers a statistical gap by 80$\times$ on average.
% That of  Huawei cloud serverless rises up to 100$\times$~\cite{socc23-huawei-cloud}.
% Consequently, for the sake of SLO guarantees, developers tend to size functions based on the worst-case, e.g., the largest working sets. 
% This, however, incurs substantial resource inefficiency.
% Production traces in Huawei cloud serverless suggests that half of deployed functions have CPU usage and memory usage merely as nearly 0.1\% and 19.5\%, respectively~\cite{socc23-huawei-cloud}.  

One promising approach for addressing such resource inefficiency is to allow for runtime resource adaptation at the request level. However, the practicality of such an approach is limited by the information barrier between the application developer and the serverless provider. Specifically, application developers do not have real-time access to runtime information necessary for per-request resource adaptation\footnote{Monitoring services like Azure Monitor~\cite{azure-monitor} and AWS CloudWatch~\cite{aws-cloudwatch} can report function runtime metrics only at one-minute intervals.}. Meanwhile, the serverless provider lacks the necessary domain knowledge of the application to fine-tune the resources without violating the SLO. 
Existing works like Kraken~\cite{socc21-kraken} and Xanadu~\cite{middleware20-xanadu} employ proactive and reactive resource scalers simultaneously to provision dynamic DAGs where only a subset of functions are invoked per request. Fifer~\cite{middleware20-fifer} and BATCH~\cite{sc20-batch} allow adjusting function sizes dynamically to achieve high resource utilization with SLO guarantee. While being effective in addressing resource inefficiency, all of them ignore the aforementioned information barrier in real-world systems, rendering them impractical for the current serverless service model.

We propose \namex, a novel runtime resource adaptation framework for serverless workflows. The goal of \namex is to achieve high resource efficiency while guaranteeing workflow SLOs. To this end, \namex adopts a late-binding approach where the developer synthesizes hints containing rules and options for resource adaptation for the serverless provider to perform runtime adaptation on their behalf following the hints passed to them. During the execution of the serverless workflow, when a function in the application DAG finishes, the serverless platform collects the execution time of that function and derives the time budget for the rest of the workflow. Based on the derived time budget, \namex adjusts the sizes of downstream functions using the hints provided by the developer which are ensured to meet the SLO requirement. 

The developer synthesizes the hints through comprehensive profiling. Different from current practice which uses P99 of function execution time to calculate the resource allocation, \namex allows the developer to explore different percentiles and obtain the corresponding resource demands as part of the hints. Such detailed hints allow the serverless platform to perform fine-grained resource adaptation, exploiting runtime information to optimize resource allocation to its maximum potential. However, sharing the detailed profiling information and letting the serverless platform search in a large space at runtime for the best resource configuration come with significant space and time overhead. \namex addresses this issue by condensing the hints while retaining their quality. 

% We propose \namex---a first-of-its-kind middle-ware fully considering the practicality when conducting adaptive resource allocation for serverless workflows.
% The goal of \namex is to achieve high resource efficiency while guaranteeing SLOs.
% To this end, \namex exploits the bilateral efforts of developers and providers.
% On developers side, \namex utilizes diverse percentile distributions to profile runtime performance variance.
% Moreover,  \namex regards percentiles as a knob for exploring higher resource efficiency without violating SLOs. 
% Notably, to keep high time-efficiency \namex regulates developers to synthesize hints---comprehensive and straightforward adaption rules in an offline manner while fully condensing them into a compact table.
% On providers side, \namex takes advantage of providers' capability in  prompt runtime information collection while relying on the table to instruct them to accomplish proper runtime adaptations.
%Based on that, providers can promptly and accurately adapt resource at runtime.

Overall, this paper makes the following contributions. After introducing the background and motivating the idea (\S\ref{sec:background}), we
\begin{itemize}
    \item  present the design of \namex---a novel runtime resource adaptation framework for serverless workflows to achieve high resource efficiency following a late-binding approach (\S\ref{sec:system-overview}),
    \item present effective algorithms for synthesizing, condensing, and utilizing hints to realize resource- and time-efficient runtime resource adaptation (\S\ref{sec:synthesizer}), 
    \item implement \namex and perform extensive experiments with two real-world serverless workflows (\S\ref{sec:evaluation}). Experiment results show that \namex  is able to to improves resource efficiency by 29.9\% and 34.7\% on average respectively, compared with the state-of-the-art serverless system, while guaranteeing latency SLOs.
\end{itemize}
\S\ref{sec:relatedwork} discusses related work and \S\ref{sec:conclusion} draws final conclusions.

\section{Background and Motivation}
\label{sec:background}

We introduce the background on serverless workload serving and motivate the use of runtime resource adaptation to address resource inefficiency in existing serverless platforms.

\subsection{Resource Inefficiency with Early Binding}
% In current serverless platforms, developers are required to specify immutable sizes for their deployed functions.
% Then, providers consider functions' runtime workloads  (e.g., concurrency)  and resource usage to scale out/in their instances.
% Moreover, due to high runtime variability, functions must size their functions for worst-case scenarios.
% This, however, incurs considerable resource inefficiency.
Current serverless workflow platforms (e.g., AWS Step Functions~\cite{aws-step-function} and Azure Durable Functions~\cite{azure-durable-function}) offer the opportunity for developers to build various applications with advanced logic like chaining, branching, and parallel execution.
These applications can be defined by JSON-based structured languages (e.g., Amazon States Language) or other programming languages.
Meanwhile, developers require to specify resource configurations, including memory size, CPU cores, and scaling options, for individual functions---an early-binding approach.
The serverless platform is responsible for monitoring the workload intensity and resource usage at runtime and scaling out/in function instances accordingly.
To account for potential runtime variability, developers must size the functions in their application workflow accounting for the worst case in order to provide SLO guarantees over the end-to-end delay of request processing, e.g., the 99th percentile (P99) of the end-to-end delay must be within a given target. 
After deployment, the function sizes become immutable. The worst case is not representative and over-shoots most of the time, leading to resource inefficiency.

To verify this claim, we conduct a data-driven analysis with a dataset from Microsoft Azure Functions~\cite{azure-dataset} to explicitly demonstrate the resource inefficiency issue. % , deriving from the worst-case based early bind.
To quantify the inefficiency, we define a metric called \emph{slack}---the margin between the actual execution time and the SLO, which is calculated as $1-l/T$ with $l$ and $T$ representing end-to-end latency and SLO, respectively.
Under certain SLO defined with P99 latency as done by existing works (e.g., \cite{osdi22-orion,mac22-wisefuse}),  we can see from Figure \ref{fig:bg:slack} that more than 60\% function invocations have slacks over 60\%.
Particularly, we analyze slacks of the top 100 most popular functions, whose invocations account for 81.6\% of the total function invocations. % (depicted in Figure~\ref{fig:bg:popular_func}) of overall invocations.
The result shows that only 20\% of the invocations of the popular functions (blue line in Figure~\ref{fig:bg:slack}) have slacks less than 40\%.
This means the majority of requests are processed faster than necessary.
Notably, in DAG-based workloads (i.e., Azure Durable Functions), the resource inefficiency further deteriorates wherein the ratio between the 95th percentile and 50th percentile is by up to three times \cite{mac22-wisefuse}.

% \begin{figure}[t!]
% \centering
% \includegraphics[width=0.25\textwidth]{./figure/motivation/Average_P99_cdf_top=100.pdf}
% \vspace{-0.3cm}
% \caption{Sufficient function slacks in production traces.}
% \label{fig:bg:slack}
% \end{figure}

\subsection{Runtime Dynamics}
\label{sec:bg:worst-case}

The resource inefficiency caused by the large slack can be mainly attributed to the over-provisioning of resources by the developer. This is to ensure that the SLO is guaranteed even in the worst case (i.e., P99). However, normal cases deviate from the worst case significantly due to runtime dynamics. 
In particular, we observe that functions face two major dynamic factors at runtime: varying working sets and inevitable performance interference. These two factors contribute significantly to the variance of the function execution time. 
% Functions face two remarkably dynamic factors at runtime: working sets and performance interference, which lead to considerable variance of execution latency.

\begin{figure*}[!t]
	\centering
	\subfloat[]{
		\includegraphics[width=0.24\textwidth]{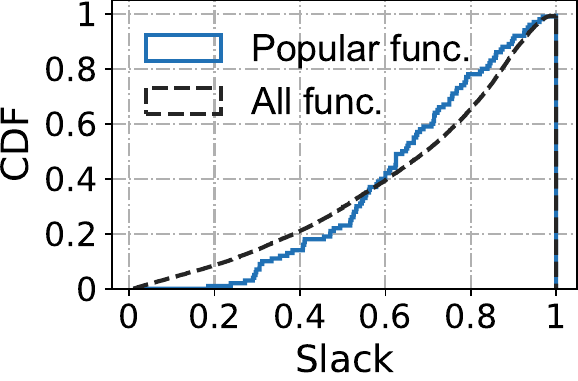}
		\label{fig:bg:slack}
	}
	\hspace{8mm}
	\subfloat[]{
		\includegraphics[width=0.25\textwidth]{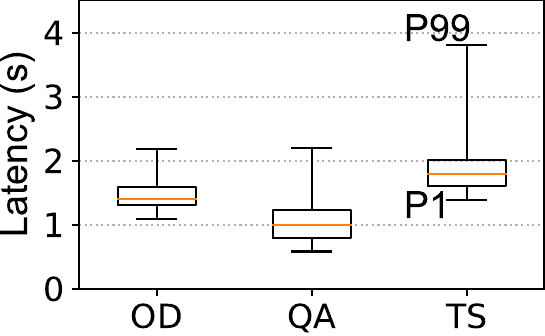}
		\label{fig:bg:ml-func-latency}
	}
	\hspace{8mm}
	\subfloat[]{
	\includegraphics[width=0.28\textwidth]{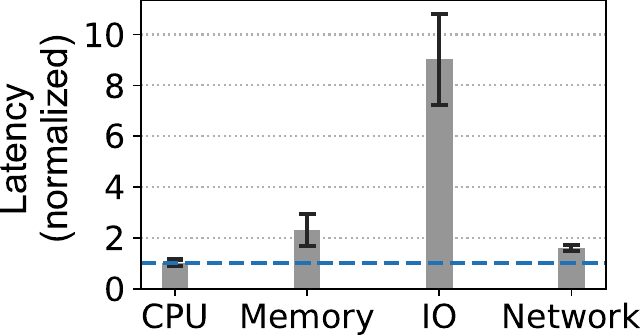}   
	\label{fig:bg:perf-inteference}
	}
	%\vspace{-0.1cm}
	\caption{(a) slacks of function invocations in production traces, (b) function latency variance caused by varying input worksets for functions object detection (OD), question answering (QA), and and text-to-speech (TS), respectively,
 (c) performance interference attributed to co-location of homogeneous function with different dominant resource demands.}
 %\vspace{-0.4cm}
\end{figure*}

%'ml-analyze':{'text-to-speech': 'text-to-speech', 'question-answer': 'question answer',
%                      'object-detection': 'object detection'
\textbf{\textit{Varying working sets.}} 
The working set, i.e., input data like videos, audios, and texts, can have varying sizes.
Taking Microsoft Azure Function Blobs (storage service) as an example, their data size difference can be as high as nine orders of magnitude~\cite{azure-function-blob}.
Such a large difference results in substantial variance of the execution time even for the same function~\cite{socc21-faast,eurosys21-ofc}.
Specifically, we measure the execution time of three functions under different working sets (detailed in \S\ref{exp:setup}).
Figure~\ref{fig:bg:ml-func-latency} illustrates the results, where we can observe a variance of up to 3.8 times in function execution caused by varying working set sizes.

% \begin{figure}[t!]
% \centering
% \includegraphics[width=0.25\textwidth]{././figure/motivation/function-latency-ml-analyze-varying-worksets.pdf}
% \vspace{-0.3cm}
% \caption{Function latency variance caused by varying input worksets}
% \label{fig:bg:ml-func-latency}
% \end{figure}	

\textbf{\textit{Performance interference.}}
% On the other hand, function deployment, which decides when and where to deploy functions, is completely undertaken by providers.
For simplicity and security, commercial serverless platforms, such as Alibaba Function Compute, Microsoft Azure, and AWS Lambda, exclusively deploy function instances belonging to the same tenant, or even belonging to the same function, in the same virtual machine~\cite{socc22-owl,atc18-peek-bench}.
For example, the empirical study in~\cite{socc22-owl} shows that in Alibaba Function Compute 65\% of the virtual machines exclusively deploy instances of the same function.
This co-location of homogeneous function instances, however, can incur severe resource contention on the same resource dimensions, particularly for network bandwidth and memory bandwidth of virtual machines~\cite{sc21-gsight,micro19-faaSprofiler,socc22-owl,atc18-peek-bench}.
To verify this observation, we use a virtual machine to run a function increasing the number of co-located instances from one to six while measuring the execution time of four different functions with resource dominance on different dimensions namely computing, I/O, network, and memory, respectively (detailed in \S\ref{exp:setup}). 
As shown in Figure~\ref{fig:bg:perf-inteference}, the co-location of homogeneous functions leads to substantial resource contention and performance interference, prolonging the function execution time up to 8.1 times. The performance interference is often hard to model and predict.

% this co-residency results in substantial increase of execution latency by up to 8.1 times,leading to considerable variance in function execution time.
% when compared with that with concurrency as one.

%for CPU-, IO-, network- and memory-intensive functions as the concurrency rises from one to six.
%Figure shows that significant performance interference can be observed, . 
%compared with the inclusive deployment (concurrency as one), 
% this exclusive deployment (gray bar) results in substantial increase of execution latency by up to 8.1$\times$ for CPU-, IO-, network- and memory-intensive functions as the concurrency rises from one to six.

% this exclusive deployment (gray bar) results in substantial increase of execution latency by up to 8.1$\times$ for CPU-, IO-, network- and memory-intensive functions as the concurrency rises from one to six.
% As depicted in Figure~\ref{fig:bg:concurrent_latency}, with the concurrency rising  from one to six,  the exclusive deployment results in substantial increase of execution latency by up to 8.1$\times$.
% This significantly magnifies execution latency variance.

% \begin{figure}[t!]
% \centering
% \includegraphics[width=0.25\textwidth]{./figure/motivation/coresident-perf.pdf}
% \vspace{-0.3cm}
% \caption{Performance interference attributed to co-residency of homogeneous function.}
% \label{fig:bg:perf-inteference}
% \end{figure}

\subsection{Runtime Resource Adaptation}
\label{sec:bg:adaptive-allocation}
To tackle the aforementioned resource inefficiency issue, we can adopt a late-binding approach through \emph{runtime resource adaptation}, which resizes functions on the fly based on runtime information (e.g., function slacks), achieving higher resource efficiency without violating SLO. For example, given a workflow as a chain of functions, the resource allocation of the downstream functions can be adjusted when the first function finishes execution. This way, the slack from the first function can be exploited to optimize resource efficiency. 

The idea sounds straightforward and has been considered in some existing works \cite{infocom22-stepconf,middleware20-fifer,socc21-llama,socc21-kraken,middleware20-xanadu}.
However, most of these works make an unrealistic assumption that either the developer performs the adaptation decision with access to runtime information or the serverless platform provider performs the adaptation with domain knowledge of the application workflow. These assumptions render these solutions impractical to deploy in real-world serverless systems. The information barrier between the developer and the provider calls for a new solution. 

We identify the following challenges and opportunities for a full-fledged design for runtime resource adaptation. 

\textbf{\textit{Skewed function execution time distribution.}} 
Resource allocation for a serverless workflow is typically done by leveraging performance profiles of all the functions in the workflow. 
During the offline profiling, the execution time distribution for each function is first obtained by running the function with a variety of sample inputs under different resource conditions. Then, given a time budget, existing approaches typically use P99 of the function execution time as a target and calculate the corresponding resource demands. However, due to the high runtime variability, the distribution of the function execution time is highly skewed where the difference between P50 and P99 can be as high as 100 times~\cite{socc23-huawei-cloud}. This means that if only the function execution time at a single percentile (P50 or P99) is used for resource allocation, there will be significant resource under-provisioning and over-provisioning for most requests at runtime. To address this issue, our idea is to allow for the exploration of the function execution time at diverse percentiles during resource allocation.

% It is a prerequisite to profile execution latency for adaptive resource allocation.  
% As aforementioned, owing to a variety of unexpected runtime dynamics,  execution latency demonstrates skewed distributions, by up to 100$\times$ between 99\% percentile and 50\% percentile on Huawei cloud serverless~\cite{socc23-huawei-cloud} .
% This makes the current a single statistic (e.g., mean) or 99\% percentile distribution based profiling suffer significant under- and over-estimation.
% To fix this issue, our insight is to \textit{introduce more diverse percentiles to profile execution latency}. 

\textbf{\textit{Dependencies of adaptation decisions.}}
As the function execution progresses, a sub-workflow will be generated by removing the finished function(s) from the workflow. Within each sub-workflow, the resource adaptation decisions for remaining functions are dependent on each other due to the constraint imposed by the end-to-end latency SLO. For example, under-provisioning a function will result in a reduction of the time budget for executing its downstream functions, thus calling for more resources for these downstream functions to avoid SLO violations. Meanwhile, the selection of the percentile for the execution time of each function dictates resource-latency tradeoff for that function. For example, a higher percentile means that more resources will be allocated to ensure that more requests processed by the function will finish within the given time budget. On the contrary, a lower percentile means that more requests will risk SLO violation, but at the benefits of reduced resource consumption. To address such complex dependencies, we propose the following ideas: (1) We introduce two metrics (i.e., the timeout metric and the resilience metric detailed in \S\ref{sec:profilier}) to balance the resource adaptation decisions of the head function of the current sub-workflow and those of the remaining downstream functions. These metrics help us connect the decision making across sub-workflows and avoids sub-optimal adaptation decisions in each sub-workflow. 
(2) We explore lower percentiles for the head function and a high percentile (i.e., P99) for other functions in each sub-workflow. Using lower percentiles maximizes the opportunity for resource optimization since any over-time execution of the head function can later be compensated by resource adaptation in the next round. The high percentile ensures that the resource adaptation is not too radical to cause SLO violations.

\textbf{\textit{Tight resource adaptation window.}}
Runtime resource adaptation requires to calculate a new resource allocation decision for the remaining sub-workflow immediately when a function finishes execution. Since serverless functions are typically short-lived (less than 1s on average)~\cite{atc18-peek-bench,socc22-owl,atc20-serverless-in-the-wild,socc23-huawei-cloud}, the window for resource adaptation is quite tight. Assuming the serverless platform will perform the runtime adaptation on behalf of the developer since the platform has access to full runtime information, the resource adaptation decision making should be fast without involving complex calculations and logic or exploring a large space. As discussed before, the serverless platform provider does not have domain knowledge of the serverless workflow. Hence, the developer must pass the necessary information to the serverless platform for runtime adaptation decision making. Our idea is to let the developer synthesize critical hints containing resource allocation rules and options, which the serverless platform provider utilizes to perform runtime resource adaptation. The hints should be highly condensed so the serverless platform can make adaptation decisions quickly enough.

% Apart from highly varying execution performance, serverless functions are also short-living (less than 1s on average)~\cite{atc18-peek-bench,socc22-owl,atc20-serverless-in-the-wild,socc23-huawei-cloud}, so is the window for adaptive allocation. 
% This variance and volatility calls for a well-preparation of hints for all possible runtime situations while promising them compact and straightforward enough for providers to easily take action.

% Here, our insight is to \emph{holistically synthesize hints in an offline manner, and then utilize the discreteness of adaptive allocation in both decision-making and decision-executing (detailed in~\S~\ref{sec:synthesizer:condense}) to fully condense the hints.
% Finally, hints are warped into a simple and compact table.
% Base on that, providers can accomplish the runtime adaption promptly and properly}.

\begin{figure}[t!]
\centering
\includegraphics[width=0.45\textwidth]{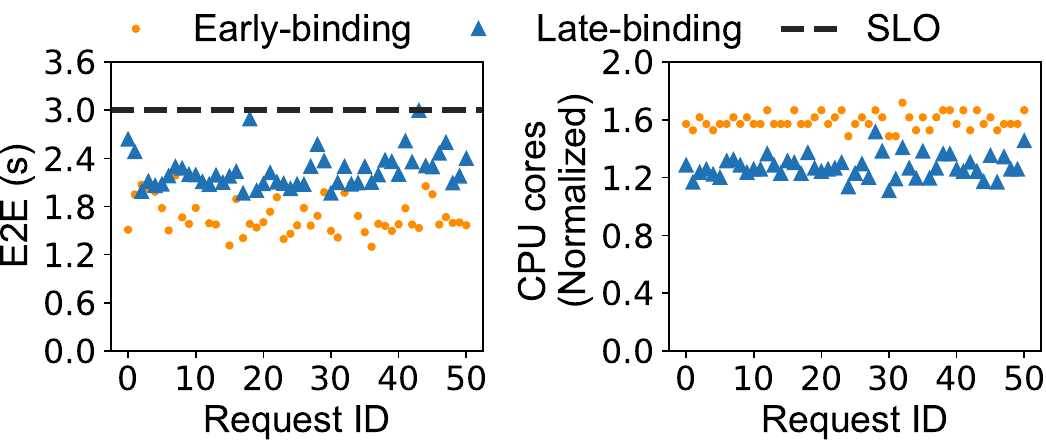}
%\vspace{-0.1cm}
\caption{Performance comparison between early-binding (left)~\cite{eurosys19-grandslam} and late-binding (runtime resource adaptation), where the CPU consumption (right) is normalized by the optimal obtained with exhaustive search.} 
%\vspace{-0.3cm}
\label{fig:bg:size}
\end{figure}

To demonstrate the potential of runtime resource adaptation incorporating all the above ideas, we take a real-world serverless workflow (explained in \S\ref{exp:setup}) as an example, and evaluate its end-to-end latency (denoted by E2E) and resource consumption (CPU cores).
As illustrated in Figure~\ref{fig:bg:size}, the late-binding (blue triangle) reduces the resource consumption by up to 42.2\% compared with existing early-binding solutions (orange circle), while ensuring SLO guarantees. This highlights the significant gains from runtime resource adaptation.

\section{\namex System Design}
\label{sec:system-overview}
We present \namex---a novel resource adaptation framework for serving serverless workflows.
The goal of \namex is to maximize resource efficiency while limiting SLO violations. \namex achieves this goal via a bilaterally engaged approach to combat the information barrier between the application developer and the serverless platform provider. 

\begin{figure*}[t!]
\centering
\includegraphics[width=0.85\textwidth]{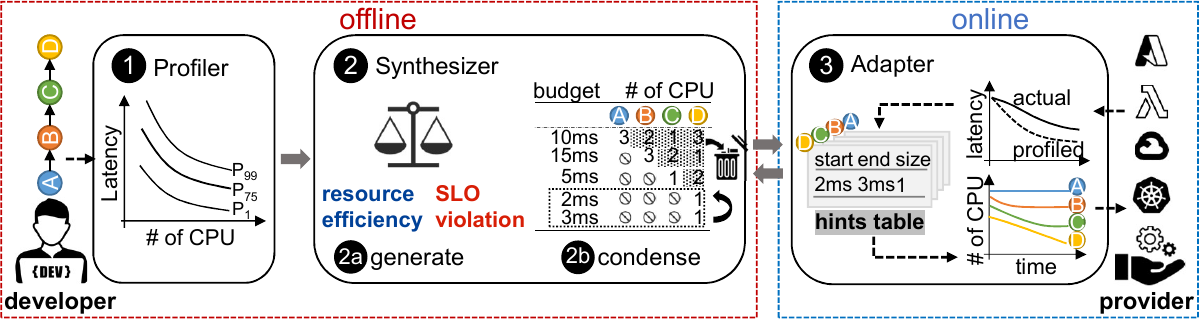}
%\vspace{-0.1cm}
\caption{An overview of the system architecture of \namex. The proposed runtime resource adaptation framework bilaterally engages the application developer and the serverless platform provider, where the developer is responsible for the offline part while the provider is responsible for the online part.}
%\vspace{-0.4cm}
\label{fig:system-overview}
\end{figure*}

\subsection{Overview}
\label{sec:system-overview:fig}
Figure~\ref{fig:system-overview} depicts an overview of the system architecture. 
\namex consists of three core components: \emph{profiler}, \emph{synthesizer}, and \emph{adapter}. 
Specifically, the profiler and synthesizer are deployed on the developer side, which are run offline, while the adapter runs online on the provider side at runtime.

The general procedure of \namex is as follows: First, the profiler interacts with the developer to collect the domain knowledge of the application, such as the workflow structure, constitutional functions execution time under varying CPU cores and concurrency settings (i.e., batch sizes), and SLO requirements.
Afterwards, the profiler extracts functions' execution time distribution (to be used as the profiles) from the collected data using different percentiles.
Then, the synthesizer takes the profiles and generates the hints table, which contains rules and options for runtime resource adaptation.
This table is submitted to the adapter on the serverless platform.
During the execution of the serverless workflow, when a function finishes, the serverless platform collects the execution time of that function and derives the time budget for the rest of the workflow.
This derived time budget is reported to the adapter, which then searches in the received hints table and notifies the platform about the adaptation decision for downstream functions.
%while relying on the hints table to adapt function sizes accordingly. 
In addition, the adapter plays the role as supervisor who carefully monitors the number of table hit/miss rates.
If the miss rate exceeds a predefined threshold, the adapter sends feedback to the developer.

Note that the developer and the provider do not generally require online, continuous interaction.
The coordination happens mostly only at the beginning of workflow deployment. 
It is expected that the submitted hints will be effective throughout the execution of the workflow. 
This is because the hints table contains fine-grained entries for time budgets produced by a comprehensive exploration of the synthesizer (detailed in \S\ref{sec:synthesizer:generate}). 
In very rare cases where hints table misses are severe (i.e., the miss rate exceeds a given threshold), the adapter notifies the developer and proposes re-triggering the profiler and synthesizer to regenerate the hints table. This regeneration process is done \emph{asynchronously} while workflow execution is still in progress, albeit with sub-optimal adaptation decisions from the adapter (explained in \S{\ref{sec:adapter}}). 

\jing{Janus performs per-workflow resource adaptation that restricts the exploitation of runtime slack within the same workflow. While this design may miss some cross-workflow optimization opportunities, it allows Janus to easily support complex scenarios, such as those involving highly parallel workflows. In a multi-user scenario, the hints are managed separately for each tenant and each workflow.}

\subsection{Profiler}
\label{sec:profilier}
% Here,  we discuss how to profile function execution latency.
% Based on that, two metrics are proposed as a preparation for synthesizing hints.
% %\subsection{Diverse percentiles based profiling}
% \textit{Diverse percentiles based profiling}
The profiler is responsible for collecting the execution time of functions under varying resources (i.e., CPU cores) and concurrency levels (i.e., batch sizes) while extracting execution time distribution by using different percentiles. 
The percentiles can be configured based on SLO requirements.
By default, we follow the widely used approach of meeting the end-to-end latency target at the 99th percentile (P99) as latency SLO~\cite{osdi22-orion,mac22-wisefuse}.
Therefore, we use percentiles ranging from 1\% to 99\% with a step of 5\% and the latency profiling of functions is done between P1 and P99. Latency numbers out of the P1-P99 range are not accounted for by Janus for optimization. 
Janus can accommodate more stringent SLO targets (e.g., at P99.9) by instructing the profiler and synthesizer to use higher percentiles (P99.9). 
% but at the cost of reduced overall resource efficiency. 

% As explained in \S{\ref{sec:bg:worst-case}}, function execution time is variable and forms a skewed distribution. 
% This renders existing resource allocation approaches that depend on a single percentile (e.g., the 99th) on such a distribution inadequate (i.e., either under-provisioning or over-provisioning) for a majority of function invocations.
% To tackle this issue, for any given batch size we introduce diverse percentiles on the function execution time distribution to calculate the resource demand, which is expressed as $L(p,k)$, with CPU cores and percentiles denoted as $k$ and $p$, respectively.

The diversity in percentiles brings more opportunities to achieve higher resource efficiency but comes at a higher risk of SLO violations.
Specifically, when setting percentiles lower than 99\%, it may cause under-estimation of function execution time, making functions prone to over-time execution, i.e., their actual execution time exceeds the profiled execution time.
To quantify the degree of potential over-time execution, we propose a metric called \emph{timeout}, which is expressed as
\begin{eqnarray}
     D(p,k) = L(99,k)-L(p,k),
\end{eqnarray}
where $L(p,k)$ represents the profiled execution time, with percentiles and CPU cores denoted as $k$ and $p$, respectively.
For preventing SLO violations, \namex must provision more processing resources for downstream functions to absorb such timeouts.
To this end, we propose a metric called \emph{resilience} to quantify the absorption capability, which is expressed as 
\begin{eqnarray}
    R(p,k)= L(p,K_{max})-L(p,k),
\end{eqnarray}
where $K_{max}$ denotes the maximum available resources.
Any timeout must be restricted within the upper bound of resilience, such that guaranteeing the SLO is still possible.

\subsection{Synthesizer}
% The synthesizer provides the intelligence of the system by generating and condensing hints in the form of a table.
% The goal of the synthesizer is to promise its generated hints with maximum resource efficiency under given time budgets.
% To this end, the synthesizer introduces percentiles as a knob to exploit a larger optimization space.
% Meanwhile, for meeting the time budget requirements, the synthesizer strictly restricts its exploration within a ``safety zone", where timeout does not exceed the resilience (explained in \S\ref{sec:synthesizer:generate}).

The synthesizer provides the intelligence of the system by generating and condensing hints in the form of a table.
The goal of the synthesizer is to produce hints with high hit rates and maximum resource efficiency.  
To this end, the synthesizer evaluates potential time budgets across a broad range considering achievable execution time of individual functions.
Based on that, the synthesizer explores diverse percentiles for functions to enhance their resource efficiency.
Moreover, the synthesizer leverages \emph{timeout} and \emph{resilience}---metrics to quantify the risk of SLO violations as detailed in \S\ref{sec:profilier}---to regulate the above exploration, aiming to provide SLO compliance.
%With that, it uses percentile as a knob to exploit a larger optimization space.
%In addition, the synthesizer restricts its exploration within a ``safety zone", where timeout does not exceed the resilience, for meeting the time budget requirements (explained in \S\ref{sec:synthesizer:generate}).

On the other hand, to keep hints tables' efficiency in both space and searching, the synthesizer makes full use of the discreteness in resource adaptation 
to condense the generated hints (detailed in \S\ref{sec:synthesizer:condense}).
Finally, the synthesizer provides a highly compact hints table with three simple fields: \textit{start}, \textit{end}, and \textit{size}.
This means any workflow with their time budget between \textit{start} and \textit{end} should be provisioned with resource amounts as \textit{size}, which can ensure the maximum resource efficiency without violating the available time budget.

% \jing{To promise hints tables with high hit rates, the synthesizer evaluates potential time budgets across a broad range.
% Specifically, it determines the upper and lower bounds of considered time budgets based on the achievable execution time of individual functions.
% Furthermore, time budgets are finely divided down to as little as 1 ms .
% }

% \jing{To keep hints tables' efficiency in both space and searching, the synthesizer makes full use of the discreteness in resource adaptation 
% to condense the generated hints (detailed in \S\ref{sec:synthesizer:condense}).}
% Finally, the synthesizer provides a highly compact hints table with three simple fields: \textit{start}, \textit{end}, and \textit{size}.
% This means any workflow with their time budget between \textit{start} and \textit{end} should be provisioned with resource amounts as \textit{size}, which can ensure the maximum resource efficiency for the (sub-)workflow without violating the available time budgets.

\subsection{Adapter}
\label{sec:adapter}
After a function in the workflow finishes, the adapter derives the available time budget for the remaining functions, and searches the hints table accordingly to figure out the appropriate resource allocation, such that the required time budget can be met with the minimum resource consumption.
If the above search results in a miss possibly due to unexpected runtime dynamics (detailed in~\S\ref{sec:bg:worst-case}), the adapter will scale functions up to the maximum available resources, to prevent SLO violations.
Afterwards, the adapter notifies the platform about the adaptation decision. 
\jing{This highly streamlined decision-making process enhances Janus's scalability.
}

On the other hand, the adapter continuously counts the hits and misses during hint table searches. 
In rare cases where the miss rate exceeds a predefined threshold, it assumes that the execution time distribution may have changed. 
In that case, the adapter notifies the developer and suggests triggering the profiler and synthesizer to regenerate hints tables asynchronously.
This asynchronous regeneration can strike the trade-off between resource- and time-efficiency in adaptation.

% \begin{algorithm}[!t]
%     \caption{Adapt resource online \label{alg:janus:adapt}}
%     \LinesNumbered    
%     \KwIn{$\mathbf{F}= \left\langle f_1,\dots,f_N \right\rangle$: (sub-)workflow}
%     \KwIn{$\mathbf{U}, T$: hints table and time budget/slack}
%     \KwOut{$k$: CPU cores for head function}
% 	\SetKwFunction{FMain}{\texttt{adapt}}
%     \SetKwProg{Fn}{Function}{:}{}
    
%     \Fn{\FMain{$\mathbf{F},\mathbf{U}, T$}}{
%     $ i \leftarrow \mathbf{U}.$\texttt{index}($T$)\\
%     \If{$i=\emptyset$}
%     {
%     $\mathbf{K} \leftarrow $\texttt{generate}($\mathbf{F},T, \emptyset$) \\
%     $\mathbf{U} \leftarrow $\texttt{condense}($\{\left \langle T, \mathbf{K}\right\rangle\}$)\\
%     }
%     \Else{
%     $\mathbf{K} \leftarrow \mathbf{U}[i]$
%     }
%     \text{return}~$\mathbf{K}[0]$
%     }

% \end{algorithm}
%\input{section/04-profiler}
\section{Synthesizer}
\label{sec:synthesizer}

We now elaborate on the workings of the synthesizer. 
The hints synthesis process consists of two steps: hints generation and hints condensing. 

% We now discuss the two steps synthesizer takes, namely hints generating and condensing.
% Profiler prepares raw material/ingredient for synthesizer to generate hints, i.e., concise and straightforward rules, to guide providers to conduct adaptive allocation at runtime.
% For maintaining high accuracy, synthesizer requires to prepare a specific hint for each given slack.
% Yet, due to the variance and volatility of slacks, \namex may generate overwhelming hints, which not only incurs extra storage resource consumption but also  hurts the time-efficiency of adaptive allocations.
% To fix this issue, \namex effectively condenses the hints.
\subsection{Hints Generation}
\label{sec:synthesizer:generate}
To generate hints tables with high hit rates and high resource efficiency, the synthesizer requires a twofold effort.
First, it must explore all potential runtime time budgets for individual sub-workflows.
Second, the synthesizer needs to balance the trade-off between higher resource efficiency and the risk of SLO violation.
To this end, we reveal the following insights.

\mypara{Insight-1: Broad time budget range.}
The time budgets are calculated based on all possibilities between the 1st and 99th percentile (P1-P99) of the function execution time under a wide range of resource allocations, aiming to achieve high hit rates.
The range of time budgets therefore are formulated as
\begin{eqnarray}
    T_{min}= \sum_{i=1}^NL_i(1,K_{max}),
    T_{max}= \sum_{i=1}^NL_i(99,K_{min}),
\end{eqnarray}
where $K_{min}$/$K_{max}$ represents the minimum/maximum available resources,
and $N$ represents the numbers of functions in the given sub-workflows. Within this range, the synthesizer explores the potential time budgets with finer granularity in milliseconds, while evaluating their corresponding resource allocation.
The synthesizer can also be configured with higher percentiles (e.g., P99.9) to meet more stringent SLO targets.

\textbf{\textit{Insight-2: Moderate percentile exploration.}}
Diverse percentiles provide more opportunities for resource optimization, but come with exponentially higher time complexity for runtime resource adaptation.
Here, our insight is to only open percentile exploration for the head function of the current sub-workflow while fixing other functions with P99.
This moderate percentile exploration benefits the synthesizer with higher resource efficiency, derived from its attempt at lower percentiles for the head function.
Meanwhile, it effectively reduces the search space for non-head functions, allowing the synthesizer to achieve high time efficiency.

\textbf{\textit{Insight-3: Resilience-aware.}}
Despite the potential of higher resource efficiency, diverse percentile exploration may put functions at the risk of timeouts, making workflows prone to SLO violations.
To address this shortcoming, the synthesizer strictly restricts the timeout within the resilience (the achievable reduction in function execution time by scaling resource up to the maximum possible).
Within this ``safety zone", the synthesizer tries its best to maximize resource efficiency.

\mypara{Insight-4: Heavier head.}
As explained in~\S\ref{sec:bg:adaptive-allocation}, facing substantial variability of execution performance, runtime resource adaptation requires to carry out (head) function by (head) function, so as to keep its high accuracy.
This, however, may lead to sub-optimal decisions due to the mismatch between the local objective and the global objective.
Specifically, the local objective is to maximize the sub-workflow's resource efficiency, while the global objective is to maximize the whole workflow's resource efficiency.
The whole efficiency is determined by that of each sub-workflow's head function, rather than that of sub-workflows.
To address this issue, the synthesizer magnifies the local objective's weight for head functions, aiming to calibrate for the mismatch.

As for how to set the weight, our insight is to increase the weight when facing loose SLOs, and vice versa.
This is because loose SLOs indicate lower resource requirements, which brings about higher resilience (depicted in Figure~\ref{fig:exp:resilience:cores}).
Increasing the weight can better utilize this higher resilience to explore lower percentiles, such that the workflow achieves higher resource efficiency with SLO guarantees.

Hints demonstrates explicit resource allocation that can ensure the sub-workflow with its maximum resource efficiency, i.e., the minimum resource consumption, under given time budgets.
This problem thus is formulated as follows:
% \begin{eqnarray}
% 	\min && W\cdot k_1+p \cdot \sum_{i=2}^{N}k_i +(1-p)\cdot (N-1)\cdot K_M \label{eq:hints:obj}\\
% 	\text{subject to} &&
%  %percentile latency
%  L(p,k_1)+\sum_{i=2}^{N}L(99,k_i) \leq T, \label{eq:hints:time-budget}\\
%  %timeouts and resilience
% && D(p,k_1) \leq \sum_{i=2}^{N}R(99,k_i), \label{eq:hints:timeout-resilience}\\
% &&  1 ~\leq p ~\leq 99,~p \in \mathbb{Z},\\
% && 0 ~< k_i ~\leq K_{max},~k_i \in \mathbb{R}, ~\forall i.
% \end{eqnarray}
\begin{eqnarray}
	\min && W k_1+p  \sum_{i=2}^{N}k_i +(1-p) (N-1) K_{max} \label{eq:hints:obj}\\
	\text{subject to} &&
 %percentile latency
 L_1(p,k_1)+\sum_{i=2}^{N}L_i(99,k_i) \leq T, \label{eq:hints:time-budget}\\
 %timeouts and resilience
&& D_1(p,k_1) \leq \sum_{i=2}^{N}R_i(99,k_i), \label{eq:hints:timeout-resilience}\\
&&  1 \leq p \leq 99,~p \in \mathbb{Z},\\
&& K_{min}\leq k_i \leq K_{max},~k_i \in \mathbb{R}, ~\forall i.
\end{eqnarray}
where $W$ is the weight for the head function (Insight-4), and $T$ and $N$ denote the time budget and the number of functions in the sub-workflow, respectively.
%$K_{max}$ represent the maximum available CPU cores.
Notably, only the head function can explore lower percentile $p$ (Insight-2).
Equation~\ref{eq:hints:obj} expresses the sub-workflow's expected resource consumption.
Specifically, $\sum_{i=2}^{N}k_i$ and $(N-1)K_{max}$ denote non-head functions' resource requirement without and with the head function's timeout, the probability of which is $p$ and $1-p$, respectively.
Equation~\ref{eq:hints:time-budget} ensures the sub-workflow's execution latency within the time budget.
Equation~\ref{eq:hints:timeout-resilience} restricts that the possible timeout of the head function can not exceed the total resilience of downstream functions (Insight-3).

\begin{algorithm}[!t]
\small
\caption{Offline hints generation\label{alg:generate}}
 	\LinesNumbered    
    \KwIn{$\mathbf{F}= \left\langle f_1,\dots,f_N \right\rangle$: (sub-)workflow}
    \KwIn{$[T_{min},T_{max}]$: time budget range}
    \KwIn{$W,\mathbf{P}$: weight and candidate percentiles  for head function $f_1$}
	%\KwIn{$\left\langle T, \mathbf{H} \right\rangle$: time budget and hints table.}
    \KwOut{$\mathbf{H}=\{\left\langle  t, \{ k_1,\dots,k_N \} \right\rangle\}$: functions' provisioned CPU cores under given time budget $t$, i.e., hints table
	}
    $\mathbf{H} \leftarrow \emptyset$, $\mathbf{P} \leftarrow \emptyset$   \\
    \ForEach{$t \in [T_{min},T_{max}]$}
    {
    $\mathbf{H} \leftarrow \mathbf{H} \cup \{ \left \langle t, ~\texttt{generate}(\mathbf{F},t,\mathbf{P}) \right \rangle \}$ \\
    $\text{return}~\mathbf{H}$
    }
	\SetKwFunction{FMain}{\texttt{generate}}
    \SetKwProg{Fn}{Function}{:}{}
    
    \Fn{\FMain{$\mathbf{F},t,\mathbf{P}$}}{
    \If{$\left| \mathbf{F} \right| = 1$} 
    {
     \textbf{return} \texttt{min\_resource}($f_1,t$)
    }
    %$r_{min} \leftarrow \infty,~X \leftarrow \emptyset $ \\
     \If{$\mathbf{P} = \emptyset$} 
    {
     $\mathbf{P}=$\texttt{explore\_percentile}($\mathbf{F},t, K_{max}$)
    }
    $s_{min} \leftarrow \infty,~\mathbf{K} \leftarrow \emptyset $\\
    \ForEach{$p \in \mathbf{P}$}
    {
        \ForEach{$k \in [K_{min},K_{max}]$}   
        {
          $\mathbf{Z} \leftarrow $~\FMain{$\mathbf{F} \setminus f_1, t-L_1(p,k), \{P_{99}\}$}\\
          \If{$\mathbf{Z} \neq \emptyset \wedge  D(p,k) \leq \sum{R(\mathbf{Z},P_{99})}$}
          {
          $s \leftarrow W  k +p \sum{\mathbf{Z}} + (1-p)  (\left| \mathbf{F} \right| -1) K_{max}$ \\
          \If{$s \leq s_{min}$}
          {
            $s_{min} \leftarrow s, \mathbf{K} \leftarrow \{k\} \cup  \mathbf{Z}  $
          }
          }
        }            
    }
    \text{return}~$\mathbf{K}$
    }
\end{algorithm}

The algorithm for generating hints is listed in Algorithm~\ref{alg:generate}.
To ensure hints tables with high hit rates, the synthesizer explores all time budgets comprehensively (lines 2--4). 
Specifically, for a given sub-workflow $\mathbf{F}$, the synthesizer first determines the percentiles $\mathbf{P}$ that can ensure $\mathbf{F}$'s execution time below the required time budget $t$, with assuming the maximum available CPU cores for each function (lines 8--9).
Then, the synthesizer explores the resource allocation for both head and non-head functions, denoted as $k$ and $\mathbf{Z}$, under given percentile $p$. Its goal is to minimize the expected resource consumption $s$, while promising timeout $ D(p,k)$ restricted within resilience $\sum{R(\mathbf{Z},P_{99})}$ (lines 12--17).
To accelerate the generation, the synthesizer explores different percentiles concurrently.

\subsection{Hints Condensing}
\label{sec:synthesizer:condense}
The synthesizer fully utilizes the discreteness in both decision-making and decision-executing to condense hints.

\mypara{Insight-5: Repeated hints.} There are various discrete variables, such as batch sizes and CPU cores, involved in resource adaptation. 
This leads to a significant number of redundant hints that share the same adaptation decisions despite having different time budgets.

\mypara{Insight-6: Unused fields.}
The dependencies of adaptation (explained in \S\ref{sec:bg:adaptive-allocation}) compels Janus to rely solely on the fields for head functions in given hints to maintain adaptation accuracy.
Consequently, removing the fields for non-head functions helps compact the hints without compromising accuracy.
 
The algorithm for condensing hints is listed in Algorithm~\ref{alg:condense}.
Specifically, the synthesizer first sorts the given hints $\mathbf{H}$ in descending order by their time budget (line 2).
Then, it gradually fuses the hints $\mathbf{H}[l]$ that share the identical size for head function $k_1$ as shown in line 4--10.
Finally, it warps hints into a table with three fields: $T_{start}$, $T_{end}$, and $k$, indicating that the head function of the target sub-workflow should be resized to $k$ when the sub-workflow's time budgets is between $T_{start}$ and $T_{end}$.
 
 In addition, the weight for head functions impacts the decision-making.
 Thus, the synthesizer maintains individual hint tables for different weights.
 We will evaluate the effectiveness of condensing algorithm in \S\ref{exp:micro:condense}, which suggests a outstanding compression ratio without hurting accuracy. 
 
\begin{algorithm}[!t]
\small
\caption{Offline hints condensing \label{alg:condense}}
 \LinesNumbered    
    \KwIn{$\mathbf{H}= \{\left\langle t,\mathbf{K} \right\rangle$\}: raw hints table}
    \KwOut{$\mathbf{U} = [\left\langle T_{start}, T_{end}, k\right\rangle ]$: condensed hints table}
	\SetKwFunction{FMain}{\texttt{condense}}
    \SetKwProg{Fn}{Function}{:}{}
    
    \Fn{\FMain{$\mathbf{H}$}}{
    $\mathbf{H} \leftarrow$  \texttt{sort}($\mathbf{H}$)\\
    $\mathbf{U} \leftarrow \emptyset$, $q, i, j \leftarrow 0$\\
    \ForEach{$l \in [0,\left| \mathbf{H} \right |]$}
    {
        $t, \left\langle k_1,\dots, k_N \right\rangle \leftarrow \mathbf{H}[l]$ \\
        \If{$q = 0 \vee k_1 = q$}
        {
        $j \leftarrow j+1$
        }
        \Else
        {
        $\mathbf{U} \leftarrow \mathbf{U} \cup \{ \left \langle \mathbf{H}[i].t, \mathbf{H}[j].t, q \right \rangle\} $ \\
        $i, j \leftarrow l$, $q \leftarrow k_1$
        }      
    }
    \text{return}~$\mathbf{U}$
    }

\end{algorithm}

\section{Evaluation}
\label{sec:evaluation}

\subsection{Setup and Implementation}
\label{exp:setup}

\mypara{Testbed.} 
Our system uses a server equipped with Intel(R) Xeon(R) CPU E5-2678 v3 2.50GHz with 24 physical CPU cores as the local server running Ubuntu 18.04, where \namex synthesizes hints tables.
Meanwhile, we use another server equipped with Intel(R) Xeon(R) Platinum 8269CY CPU 2.50GHz with 52 physical CPU cores, running Ubuntu 18.04, as a serverless platform.
In this platform, we implement \namex into an open-source framework as Fission~\cite{fission} (V1.16) for serverless functions on Kubernetes.
We use Fission PoolManager~\cite{fission-executor} to spin up function pods, due to its excellent performance against cold starts.

\mypara{Implementation.} \namex has a frontend side and a (remote) backend side.
To facilitate seamless coordination, we implement the profiler and synthesizer as two distinct functions on the frontend side, while deploying the adapter as a service on the backend side.
Moreover, the adapter can be equipped with automatic horizontal scaling for enhancing Janus's scalability.
The frontend interacts with the developer and depends on their domain knowledge (detailed below) to synthesize hints tables.
We leverage packages \texttt{pandas.DataFrame} to represent hints tables. 
As for the backend side, we develop a lightweight server using Python Flask~\cite{python-flask}, Redis~\cite{redis}, Fission APIs~\cite{fission-cli}, and Fission HTTP trigger~\cite{fission-http-trigger}. 
The server spawns a process to trace each request's execution.
Upon completion of any function in the workflow, this process will re-evaluate the time budget for the remaining functions while accessing hints tables to decide on proper resource adaptation.
%\jing{When a hints table miss occurs, the adapter will specify the maximum available CPU cores, i.e., 3000 millicores, for the target functions to prevent SLO violations.
%Additionally, the adapter records these misses and evaluates them against a predefined threshold to decide whether to notify the developer.
%The developer are responsible for updating and resubmitting hints tables by re-triggering the profiler and synthesizer.}

\mypara{Workflows.} We evaluate the effectiveness of \namex with two real-world serverless workflows namely Intelligent Assistant (IA) and Video Analyze (VA).
Specifically, IA is a chain constituted by three functions: \textit{object detection} (OD, for short)~\cite{object-detect}, \textit{question answer} (QA)~\cite{question-answer}, and \textit{text-to-speech} (TS)~\cite{text-to-speech}, which analyzes images randomly sampled from COCO2014~\cite{coco-dataset} and answers questions sampled from SQuAD2.0~\cite{qa-dataset}; finally, the answers return in the form as audios.
VA as another workflow chain includes three functions as \textit{frame extraction} (FE)~\cite{extract-frame}, \textit{image classification} (ICL)~\cite{image-classification}, and \textit{image compression} (ICO)~\cite{image-compression}.
Its inputs are YouTube videos with identical duration and resolution, sourced from ORION~\cite{osdi22-orion}.
Additionally, the four functions in \S\ref{sec:bg:worst-case} as \textit{CPU-}, \textit{Memory-}, \textit{Network-}, and \textit{IO-intensive} conduct AES encryption~\cite{isca22-lukewarm}, data read (from a Redis based in-memory database)~\cite{middleware21-sebs}, socket communication~\cite{middleware21-sebs}, and data write (to local disks)~\cite{atc18-peek-bench}, respectively.

\mypara{Runtime dynamics.}
\jing{Our testing workloads contain runtime dynamics, encompassing varying working sets and performance interference.}
\jing{Specifically, the input data for IA, i.e., images and texts from COCO2014 and SQuAD2.0 respectively, are with varying working sets. 
The empirical study shows that the number of objects per image in COCO2014 ranges from 1 to 15~\cite{instance-per-image-coco2014}, while the number of words per text in SQuAD2.0 ranges from 35 to 641. As shown in Figure~\ref{fig:bg:ml-func-latency}, these varying sets result in significant variance in function execution time.
On the other hand, VA extracts frames from videos, followed by image classification and compression. To accelerate processing, VA implements parallelism for each function, incurring cross-function performance interference inevitably. The profiles reveal that, for the three functions in VA, the average ratio of execution time between P99 and P50 is 1.46 times, 1.56 times, and 1.37 times, respectively.}

\mypara{Domain knowledge.} 
We collect the execution time of IA's and VA's functions with respect to CPU cores, ranging from 1000 millicores to 3000 millicores with a step of 100 millicores.
After data collection, \namex adopts diverse percentiles, ranging from 1\% to 99\% with a step as 5\%, to profile execution time distribution.
To assess \namex's performance over higher loads, we additionally profile IA's execution performance over higher concurrency (i.e., batch size) as two and three.
\jing{As for VA, we only profile its performance with concurrency as one because FE and ICO cannot process frames in batch form.}
Here, we exclude memory as a knob.
This is because \namex focuses on latency-critical workflows.
Our empirical tests show that memory has no impact on execution time.

\begin{figure*}[!t]
    \centering
\includegraphics[width=0.93\textwidth]{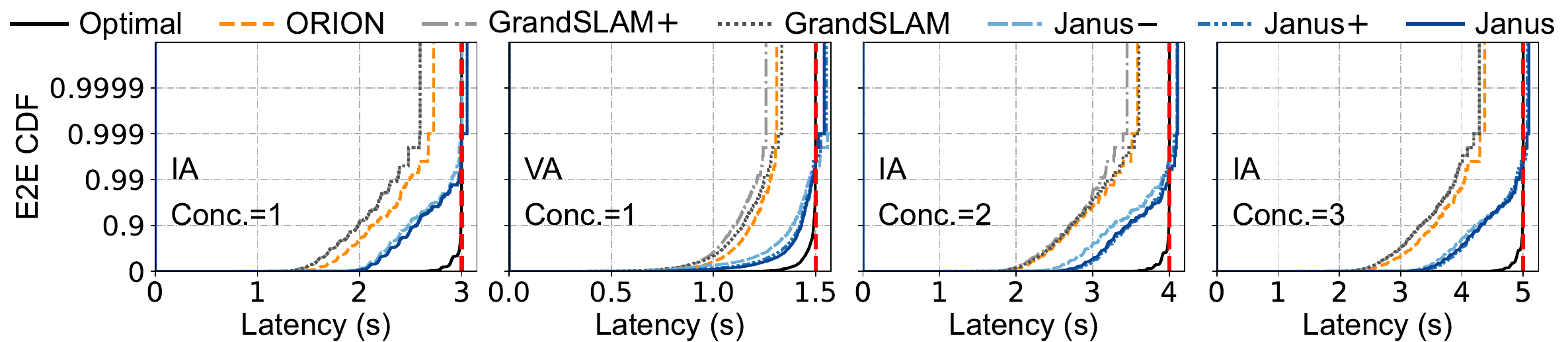}
    %\vspace{-0.1cm}
    \caption{
    End-to-end latency distribution of IA under the concurrency (i.e., batch size) as one, two and three respectively, with different SLOs (red dashed line). The concurrency of VA is limited to one due to its \jing{non-batchable} functions (i.e., FE and ICO). 
    \label{fig:exp:macro:e2e}}
    %\vspace{-0.4cm}
\end{figure*}

\mypara{Baselines.}
\namex proposes bilaterally engaged resource adaptation to provide efficient serverless workflows serving, aiming to maximize resource efficiency, i.e., minimize resource consumption, with SLO guarantees.
\jing{Here, we use three early-binding approaches and three late-binding approaches as our baselines.
The early-binding approaches include the state-of-the-art serverless workflow serving system ORION~\cite{osdi22-orion}, GrandSLAM~\cite{eurosys19-grandslam} and its enhanced version GrandSLAM$+$. 
Specifically, GrandSLAM$+$ improves GrandSLAM by removing the
latter’s constraints in identical sizes for all functions.
The late-binding approaches include Janus$-$, Janus$+$, and Optimal.
Optimal represents the best that can be achieved in any late-binding solution.}

\jing{The differences between Janus, Janus-, and Janus$+$ are as follows: Janus allows exploring diverse percentiles for the head (first) function in workflows. 
Janus$-$ disables this exploration and adopts a fixed percentile, P99. 
Janus$+$ extends the exploration to both the head function and the next-to-head function. 
In summary, compared to Janus, Janus$-$ has compromised resource efficiency due to its smaller optimization space. 
Janus$+$ can have higher resource efficiency (e.g., 0.6\% higher than Janus for IA) owing to a larger optimization space but at the expense of considerable time cost (by up to 107.2 times) in synthesizing hints (\S\ref{exp:micro:vary-perc}).}

\jing{Notably, existing late-binding approaches including Fifer, Kraken, Xanadu, and Cypress~\cite{socc22-cypress} mostly overlook the information barrier between the developer and provider, raising practicality concerns.
Additionally, as highlighted by Cypress, Fifer, Kraken, and Xanadu assume that function execution time does not have large variance, and hence they adopt mean execution time to perform runtime resource adaptation. 
However, this assumption contradicts our empirical observations from serverless production traces, which exhibit significant variance in execution time (\S\ref{sec:background}). 
Consequently, these approaches are easily prone to under provisioning and severe SLO violations. 
Thus, we exclude them as the baselines.}

% \namex proposes bilaterally engaged resource adaptation to provide efficient serverless workflows serving, aiming to maximize resource efficiency, i.e., minimize resource consumption, with SLO guarantees.
% Here, we use the state-of-the-art serverless workflow serving system ORION~\cite{osdi22-orion}, GrandSLAM~\cite{eurosys19-grandslam} and its enhanced version GrandSLAM$+$, Janus$-$, and Optimal as our baselines.
% Specifically,  GrandSLAM$+$ improves GrandSLAM by removing the latter's constraints in identical sizes for all functions.
% Janus$-$ disables head functions' exploration to diverse percentiles, and adopts a fixed percentile as P99.  
% Optimal is established through exhaustive search. 
% Specifically, we generate samples (i.e., requests) for the two workflows, respectively. For each sample we record the exact execution time of individual functions with respect to CPU cores. Moreover, the execution time is randomly extracted from the original profiling data. For each sample, Optimal searches over all possible CPU core combinations for different functions, thus aiming to minimize overall resource consumption while ensuring SLOs.

%Note that except \namex, Janus$-$ and Optimal, other systems adopt early-binding.
%Overall, we compare these systems' performance over two metrics as total resource consumption and end-to-end latency.

\mypara{Setup.}
Considering our testbed's capacity and the short-lived nature of functions~\cite{atc20-serverless-in-the-wild,socc23-huawei-cloud}, we set SLOs for IA and VA as 3s and 1.5s, respectively.
We set the weight for each function as one unless otherwise specified.
We explore \namex's performance under varying SLOs and weights in \S\ref{exp:micro:slo} and \S\ref{exp:micro:weight}, respectively.
When a hints table miss occurs, we scale functions up to 3000 millicores to prevent SLO violations.
The miss rate threshold is set as 1\% by default.
To ensure experimental results' statistical significance, we evaluate the performance of \namex and baselines over 1000 requests.

\subsection{Overall Performance}
\label{exp:macro}
\mypara{End-to-end latency distribution.} 
Figure~\ref{fig:exp:macro:e2e} shows the end-to-end latency (E2E) distribution of IA and VA under the concurrency as one, as well as that of IA under the concurrency as two and three respectively.
We observe that Janus can fulfill the SLO requirements in all cases despite relatively higher E2E.
This is because Janus aims to improve resource efficiency while meeting latency SLOs. 
Under the premise of fulfilling SLOs, Janus trades in time for resource efficiency.

\begin{figure}
    \centering
    \subfloat[]{
        \includegraphics[width=0.4\textwidth]{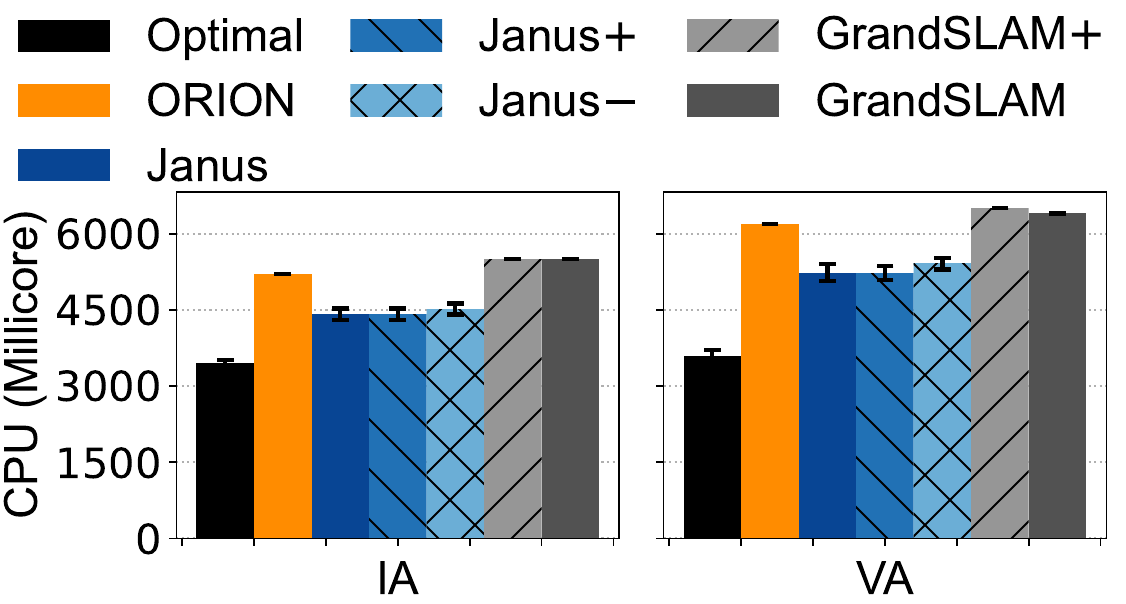}
        \label{fig:exp:macro:size}}
    \vfill
    %\vspace{-3mm}
    \subfloat[]{
    \includegraphics[width=0.41\textwidth]{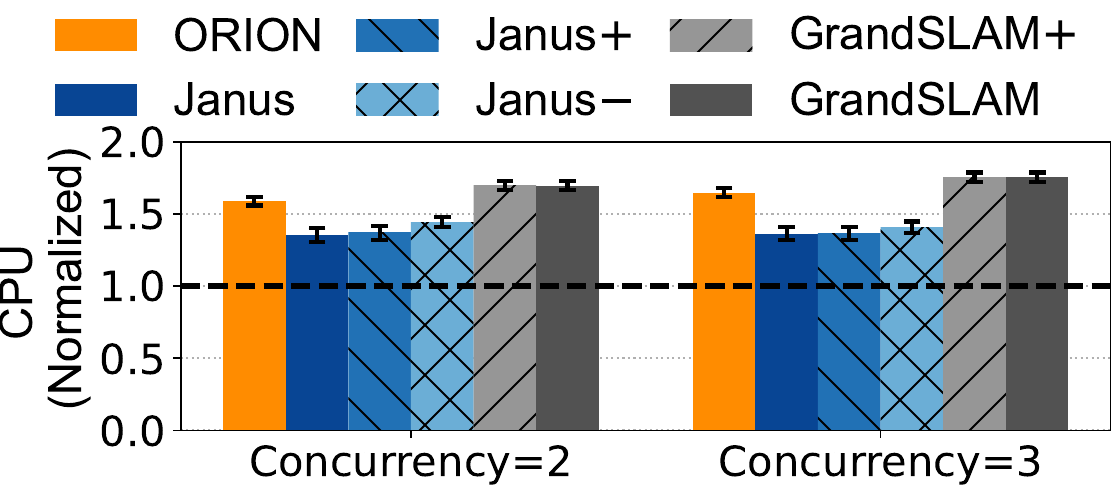}
    \label{fig:exp:macro:size-con=2-3}}
    %\vspace{-0.1cm}
    \caption{Resource consumption of (a) IA (left) and VA (right) under the concurrency as one, respectively, and of (b) IA under the concurrency as two (left) and three (right), respectively.}
   % \vspace{-0.3cm}
\end{figure}

\begin{table}
	\small  
	\centering
\caption{\label{table:exp:overall:resource} Overall resource reduction (normalized by Optimal) by \namex compared to baselines when serving IA and VA, respectively.}
	%\vspace{-0.1cm}
        \setlength{\tabcolsep}{3pt}
	\begin{tabular}{@{}l r r r r r @{}}
		\toprule
		{} &  \textbf{ORION}  & \textbf{GrandSLAM$+$} &  \textbf{GrandSLAM} &  \textbf{Janus$-$} & \textbf{Janus$+$}\\
		\midrule
		%\rule{0pt}{10pt} &\multicolumn{3}{c}{ \textbf{Plan 1}}&\multicolumn{3}{c}{\textbf{Plan 2}}\\
		IA(\%)  & 22.6 & 31.3 & 31.3 & 2.9 & 0\\
		% \midrule
		VA(\%)  & 26.9 & 35.2 & 32.4 & 4.7 & -0.2\\

		\bottomrule
	\end{tabular}\\
\label{exp:table:macro:size}
%\vspace{-0.3cm}
\end{table}

\mypara{Resource consumption.}
We compare the resource consumption of \namex and baselines when serving IA and VA given SLO as 3s and 1.5s respectively, with the concurrency as one.
Table~\ref{exp:table:macro:size} shows the average resource reduction of \namex, normalized by Optimal,  when compared with baselines, and Figure~\ref{fig:exp:macro:size} illustrates the detailed comparison.
We can see that \namex outperforms GrandSLAM$+$, GrandSLAM, and ORION significantly. 
This is because \namex fully uses slacks at runtime to improve resource efficiency. 
Compared with Janus$-$, Janus achieves further resource reduction as 2.9\% and 4.7\% for IV and VA respectively, due to Janus's exploration of lower percentiles for head functions.
\jing{Additionally, Janus incurs a negligible increase in resource consumption, i.e., 0.2\%, compared to Janus$+$.}

We also assess \namex's performance under higher loads.
Here, we increase IA's concurrency up to two and three.
For a fair comparison, we increase SLOs to 4s and 5s respectively, to promise GrandSLAM and GrandSLAM$+$ with feasible function sizes.
Figure~\ref{fig:exp:macro:size-con=2-3} shows the resource consumption normalized by Optimal. 
We find that the three early-binding systems, i.e., GrandSLAM, GrandSLAM$+$, and ORION, suffer over-allocation by up to 1.75 times.
This is because the increase of the concurrency further enlarges the runtime variability.
For example, the gap between P99 and P50 of QA (the second function in IA) increases from 2.17 times to 2.32 times on average.
This higher variability magnifies the early-binding's over-provisioning.
As a contrast, \namex relies on its runtime adaptation to capture the variance and resize functions correspondingly, thus reaping higher resource efficiency.

\subsection{Effectiveness of Moderate Percentile Exploration}
\label{exp:micro:vary-perc}
\jing{
Here, we assess the effectiveness of the moderate exploration approach (\S\ref{sec:synthesizer:generate}) adopted by Janus, which restricts the exploration of lower percentiles (below P99) to head functions within sub-workflows.
This strategy aims to balance the trade-off between the time- and resource-efficiency of resource adaptation.
We compare Janus with Janus$+$, which extends percentile exploration to both the head function and the next-to-head function.
While this expansion of percentile exploration can yield higher resource efficiency, it comes at the expense of significant time cost in synthesizing hints tables.}

\begin{figure}[t!]
    %\vspace{-0.2cm}
    \centering
	\subfloat[]{
		\includegraphics[width=0.235\textwidth]{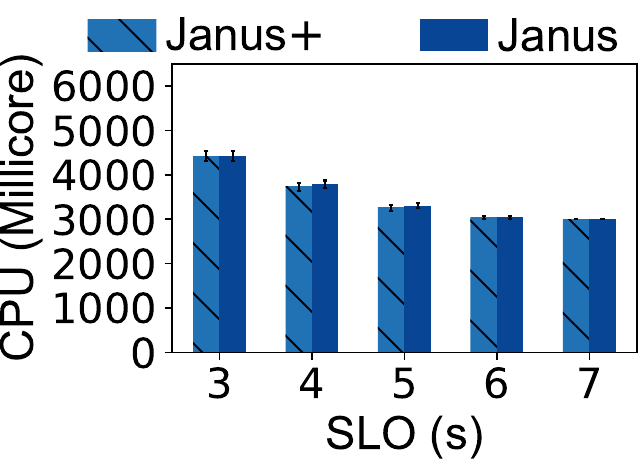}
		\label{fig:exp:vary-perc:size}}
	\hfill
	\subfloat[]{
		\includegraphics[width=0.225\textwidth]{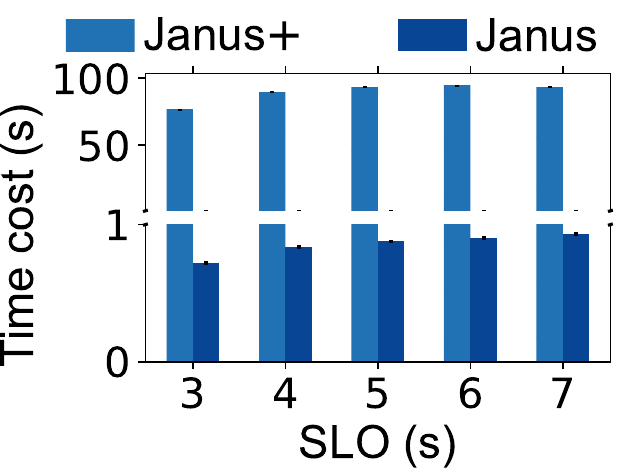}
		\label{fig:exp:vary-perc:time}}
	%\vspace{-0.1cm}
	\caption{(a) Workflow sizes and (b) time costs of Janus$+$ and \namex respectively, with SLOs ranging from 3s to 7s.}
    %\vspace{-0.3cm}
\end{figure}

Taking IA as an instance, we observe from Figure~\ref{fig:exp:vary-perc:size} that compared with Janus, Janus$+$ decreases resource consumption merely by 0.6\% on average. This means the optimization space of wider percentile exploration for IA is limited.
However, this limited reduction in resource comes at a significant time cost in synthesizing hints, by up to 107.2 times higher than \namex, as depicted in Figure~\ref{fig:exp:vary-perc:time}.

Additionally, Janus's time costs increases marginally as SLO grows.
This is because higher SLOs brings in more candidate adaptation plans.
\namex needs to efficiently evaluate these plans, and figure out the one with the minimum resource consumption, thus incurring higher time costs.
Notably,  the above time costs only happen during hints generation.
When coming to online adaptation,  its overhead is merely less than 3ms (explained in \S\ref{exp:micro:overhead}).

\begin{figure}[!t]
    %\vspace{-0.2cm}
	\subfloat[]{
    \includegraphics[width=0.23\textwidth]{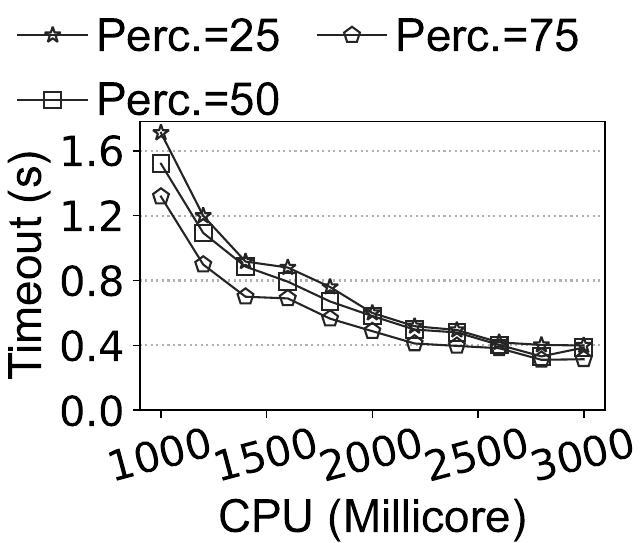}
		\label{fig:exp:timeout:ia}}
	\hfill
	\subfloat[]{
		\includegraphics[width=0.21\textwidth]{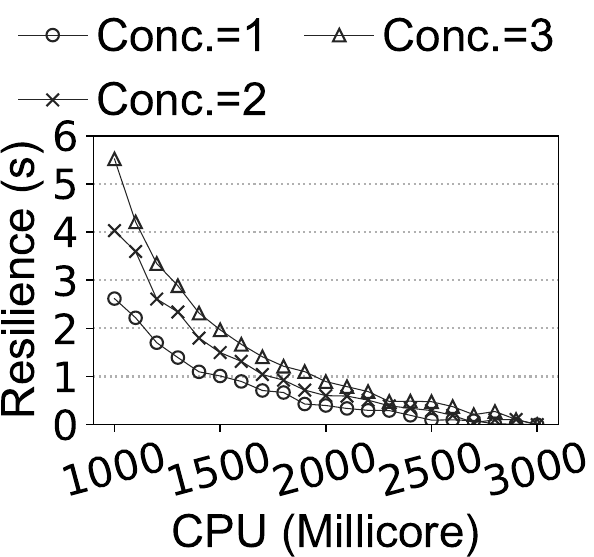}
		\label{fig:exp:resilience:cores}}
	%\vspace{-0.1cm}
	\caption{(a) Timeout and (b) resilience of TS under varying CPU cores.}
    %\vspace{-0.3cm}
\end{figure}

\subsection{Timeout and Resilience}
\label{exp:micro:timeout-resilience}
We propose timeout and resilience to quantify the risk of SLO violations (detailed in \S\ref{sec:synthesizer:generate}).
Owing to space constraints, we use TS from IA as an example, and other functions exhibit similar patterns.
We observe from Figure~\ref{fig:exp:timeout:ia} that timeout decreases as either percentiles or available CPU cores increase.
This is because additional resources enhance functions' capability to handle both runtime interference and variability of working sets\cite{socc22-owl}, thus reaping lower timeout.

As for resilience, Figure~\ref{fig:exp:resilience:cores} shows a marginal reduction as the number of provisioned CPU cores increase.
This is attributed to non-parallelizable operations within functions, leading to diminishing returns on execution time despite the addition of more resources.
Additionally, higher concurrency enhances higher resilience.
This is due to the increased computing load, which heightens functions' sensitivity to resources, thereby boosting resilience.

%Resilience varies across different functions.
%As shown in Figure~\ref{fig:exp:resilience:ia}, the resilience of TS is much higher than that of other functions.
%Similar to timeout, resilience is also highly correlated to functions' sensitivity to resources.
%Generally, the higher computing intensity functions have, higher resilience sensitivity they own.
%
%\begin{figure}
%    \centering
%    \subfloat[]{
%        \includegraphics[width=0.22\textwidth]{./figure/exp/micro/ml-analyze-func.pdf}
%        \label{fig:exp:resilience:ia}}
%    \hfill
%    \subfloat[]{
%        \includegraphics[width=0.22\textwidth]{./figure/exp/micro/video-analyze-func.pdf}
%        \label{fig:exp:resilience:va}}
%    \caption{Average resilience of (a) IA and (b) VA}
%\end{figure}

\begin{table}[!t]
        \small
	\centering
	\caption{Resource consumption and percentiles for the head function of IA with the weight as one and three, respectively.}
	%\vspace{-0.1cm}
	\begin{tabular}{@{}l r r @{}}
		\toprule
		{} &  \textbf{weight=1}  & \textbf{weight=3} \\
		\midrule
		%\rule{0pt}{10pt} &\multicolumn{3}{c}{ \textbf{Plan 1}}&\multicolumn{3}{c}{\textbf{Plan 2}}\\
		CPU (Millicore)  & 1442.9  & 1228.6    \\
		%\midrule
		Percentile (\%)  & 94.4 & 91.3  \\
		\bottomrule 
	\end{tabular}\\
	\label{table:exp:micro:weight}
\end{table}

\subsection{Impact of Weight}
\label{exp:micro:weight}
Higher weights for head functions is introduced to further improve the resource efficiency of Janus (detailed in \S\ref{sec:synthesizer:generate}).
Taking IA as an example, we evaluate its resource consumption with SLOs ranging from 4s to 10s under the weight as one and three, respectively.
The results show that when the SLO is less than 8s the moderate weight consumes less resources by 2.9\% on average.
Conversely, as the SLO becomes relaxed, the higher weight allows to further reduce resource by 1\% owing to its larger optimization space.

We also examine the impact of weights on resource consumption and percentile selection for head functions.
Table~\ref{table:exp:micro:weight} shows that \namex tends to decrease both resource allocation and percentiles under higher weights.
This is because with higher weights the objective focuses more on decreasing the size of head functions, rather than that of sub-workflows.
Lower percentiles typically indicate that fewer requests need to be completed within the specified time budget, thus requiring fewer resource consumption.
This aligns well with the objective with higher weights.
Yet, lower percentiles may expose sub-workflows at the risk of timeouts, particularly under strict SLOs.
To prevent SLO violations, non-head functions may compensate by requesting additional resources, potentially hurting the overall resource efficiency.

\subsection{Effectiveness of Hints Condensing}
\label{exp:micro:condense}

\begin{figure}[!t]
\centering
%\vspace{-0.4cm}
\includegraphics[width=0.4\textwidth]{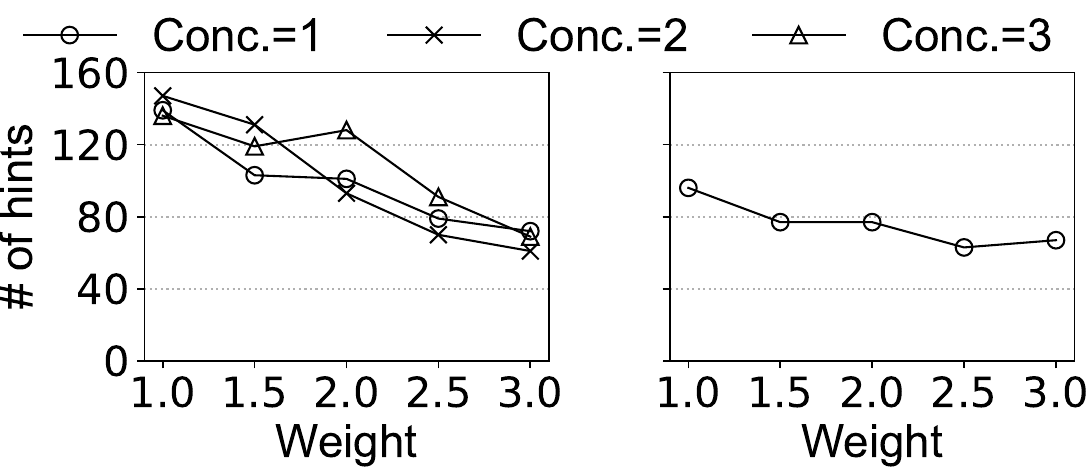}
%\vspace{-0.1cm}
\caption{Total numbers of hints synthesized for IA (left) and VA (right) under different weights.}
%\vspace{-0.3cm}
\label{fig:exp:condense:total-hints}
\end{figure}

We assess hints table sizes, i.e., numbers of hints, with and without condensing.
As explained in \S\ref{sec:synthesizer:generate} we depend on our testbed's capacity to configure the range of time budgets explored during hints generation.
Specifically, for IA the range is from 2s to 7s,  from 3s to 7s, and from 4s to 10s, with a fine-grain step of 1ms, under the three different concurrency, respectively.
For VA this range is from 1.5s to 2s with a step of 1ms.
Additionally, weight, as a hyper-parameter involving hint generation, also influences hints table sizes.
Therefore, we assess the two workflows' hints table sizes under different weights ranging from 1 to 3 with a step as 0.5, respectively.

Figure~\ref{fig:exp:condense:total-hints} illustrates the detailed number of hints for IA (left) and VA (right), respectively.
After effective condensing, the overall hints for IA and VA are less than 147 and 96 respectively, achieving compression ratios of up to 99.6\% and 98.2\%.
In addition, the size of hints tables decreases as the weight increases.
The reason is that higher weights focus more on minimizing the size of the head function, which may lead to the sub-workflow's over-allocation.
This over-allocation increases hints' applicability across different runtime conditions, thus benefiting hints tables with smaller sizes.

\begin{figure}[!t]
\centering
%\vspace{-0.4cm}
\includegraphics[width=0.4\textwidth]{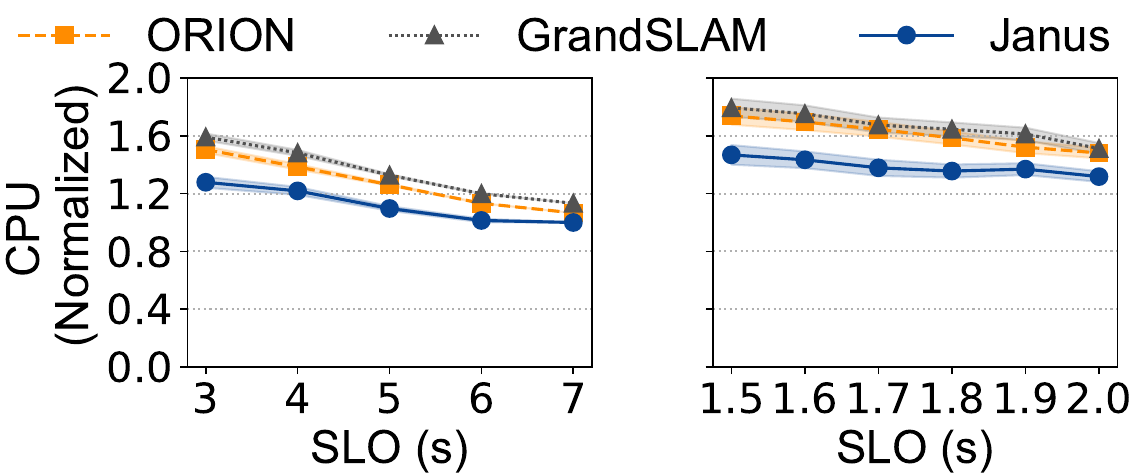}
%\vspace{-0.1cm}
\caption{IA's (left) and VA's (right) resource consumption (normalized by Optimal) under different SLOs. }
%\vspace{-0.3cm}
\label{fig:exp:slo}
\end{figure}

\subsection{Impact of SLO}
\label{exp:micro:slo}
\jing{We compare \namex's resource consumption with baselines when serving IA and VA under varying SLOs.
For clarity, we normalize the results by Optimal.
To ensure readability, we illustrate only the results of ORION, GrandSLAM, and Janus.
As shown in Figure~\ref{fig:exp:slo}, for IA \namex outperforms ORION and GrandSLAM by 16.1\% and 24.1\% on average, respectively.
In terms of VA, \namex outperforms the baselines by 22.2\% and 27.7\%, respectively.
Notably, as SLOs increase, \namex's performance gains decrease marginally.
This is due to our testbed's limitation of CPU cores, i.e., 1000 millicores at least per function, restricting \namex's further improvement.
For instance, under given SLOs as 6s and 7s, IA's resource consumption reduces to 3043.6 millicores, approaching that of Optimal (i.e., 3000 millicores).
As for other baselines, GrandSLAM$+$ exhibits performance comparable to GrandSLAM, with a marginal gap of less than 0.6\%.
Janus$+$ achieves a resource reduction of up to 1.8\% compared to Janus. Janus$-$ incurs higher resource consumption, exceeding Janus by 3.2\% and 4.3\% on average, when serving IA and VA respectively.}

%\jing{For readability, do not illustrate the results of GrandSLAM+ and Janus-/Janus+, which are provided in words.}

%\begin{figure}[h!]
%\centering
%\includegraphics[width=0.45\textwidth]{./figure/exp/micro/slo-batch=2-3.pdf}
%\vspace{-0.3cm}
%\caption{IA's resource consumption under concurrency as two (left) and three (right), respectively.}
%\label{fig:exp:slo:high-concurrency}
%\end{figure}

%Additionally, we also evaluate \namex's performance under higher workloads, i.e., concurrency as two and three respectively.%, when serving IA.
%Under given concurrency as two, \namex reduces resource consumption by 23.1\%, 30.5\%, 30.1\%, and 4.5\%, when compared with ORION, GrandSLAM, GrandSLAM$+$, and Janus$-$, respectively.
%This improvement further increases by up to 2\% as concurrency rises to three.
%Notably, when given concurrency as three, neither GrandSLAM nor GrandSLAM$+$ has feasible sizes for SLO as 4s, due to their inflexible allocation. 

\subsection{System Overhead}
\label{exp:micro:overhead}
We evaluate \namex's time cost for online resource adaptation serving IA and VA respectively, under varying SLOs from 2s to 7s with the weight as one and three, respectively.
The results show that the time cost remains under 3ms.
This suggests that Janus maintains high time-efficiency unaffected by either SLOs or weights. 

We measure the memory footprint of Janus during online adaptation and offline hints generation.
As for online adaption, Janus consumes negligible memory less than 12.1MB and 10.9MB for IA and VA, respectively.
In terms of offline hints generation, the average memory consumption is less than 12.4MB and 10.9MB for IA and VA, respectively.

\section{Related Work}
\label{sec:relatedwork}

We summarize related work covering both early-binding and late-binding approaches for serverless resource management.

% \begin{table}
%     \centering
%     \begin{tabular}{c|c|c c}
%      Name  & \textbf{Early-bind} & \multicolumn{2}{c}{\textbf{Late-bind}}\\
%      \hline
%          &  & Decision making& Runtime adaptation\\
%     \hline
%          COSE~\cite{infocom20-cose}  &  \cellcolor{lightgray}&  & \\
%          SIMPPO~\cite{socc22-simppo}  & \cellcolor{lightgray} & & \\
%          FA2~\cite{rtas22-fa2}  & & &   \\
         
%          \jing{BATCH}~\cite{sc20-batch} & &   \cellcolor{lightgray}D &\cellcolor{lightgray}D 
%     \end{tabular}
%     \caption{D and P represent developers and providers.}
%     \label{tab:my_label}
% \end{table}

\subsection{Early Binding}
COSE~\cite{infocom20-cose}, Sizeless~\cite{middleware21-sizeless}, and Parrotfish~\cite{socc23-parrotfish} adopt machine learning to learn the cost/performance of functions with respect to different sizes, and then select ``best" function sizes, such that overall costs can be minimized without violating SLOs.
FA2~\cite{rtas22-fa2} fully considers the dependency of functions within a workflow and their uncertain execution paths to periodically adapt resources, aiming to minimize resource consumption while promising SLAs.
Aquatope~\cite{asplos23-aquatope} considers runtime performance interference to decide function sizes.
ORION~\cite{osdi22-orion} and WISEFUSE~\cite{mac22-wisefuse} observe skewed function execution latency and develop a distribution-based performance modeling to provision serverless DAGs.
\jing{GrandSLAM~\cite{eurosys19-grandslam} provisions functions with fixed and identical sizes, while dynamically batching and reordering requests within each function by considering runtime slacks, aiming to achieve higher throughput without violating SLOs.}
Additionally, Morhpling~\cite{socc21-morphling} and INFaaS~\cite{socc21-llama} focus on resource auto-configuration for ML-inference specific systems.
On the other hand, there exists research from the industry that helps developers decide function sizes, such as AWS Lambda Power Tuning~\cite{lambda-tuning} and AWS Compute Optimizer~\cite{lambda-compute-optimizer}.

\subsection{Late Binding}

Cirrus~\cite{socc19-cirrus} proposes a serverless framework to boost the performance of best-efforts tasks (i.e., ML training), by integrating a client-side to monitor the remote execution while adjusting its resource allocation.
Fifer~\cite{middleware20-fifer} leverages the slacks, generated at each stage, to adjust batch sizes and scale out/in containers for higher resource utilization with SLO guarantees.
Atoll~\cite{socc21-atoll} enables proactive resource scaling as well as deadline-aware scheduling to minimize SLO violations.
BATCH~\cite{sc20-batch} fully considers serverless workload burstiness (the intensity of arrival requests) to dynamically adjust function size (memory size) and batching parameters, for the sake of minimizing monetary cost without violating SLOs.
Kraken~\cite{socc21-kraken} and Xanadu~\cite{middleware20-xanadu} employ proactive and reactive resource scalers simultaneously to provision dynamic DAG workloads, which have uncertain execution paths, aiming to minimize resource consumption without SLO violations.
Cypress~\cite{socc22-cypress} enables input size-aware request batching and resource provisioning. 
Llama~\cite{socc21-llama} focuses on auto-tuning video analytics pipelines under heterogeneous serverless environments.
Erms~\cite{asplos23-erms}, FIRM~\cite{osdi20-firm}, and Sinan~\cite{asplos21-sinan} focus on improving resource efficiency without violating SLOs, for shared microservices.
Apart from auto-scaling, there are works focusing on serverless workflows scheduling~\cite{socc20-sequoia,arxiv22-resc,ic2e22-fusionize,socc20-wukong,atc21-sonic,atc21-faastlane,asplos22-faasflow,sigcomm24-yuanrong} and over-commit~\cite{socc22-owl,socc23-golgi}.

% For example, Sequoia~\cite{socc20-sequoia} offers multiple scheduling policies available to providers/developers to alleviate issues mid-chain drops.}

% \qf{Sequoia~\cite{socc20-sequoia} optimizes workflow scheduling strategies by leveraging real-time monitoring of system and function states.}

% \qf{Sonic~\cite{atc21-sonic} proposes a layered data storage architecture to address the performance bottlenecks caused by serverless functions' interaction. Faastlane~\cite{atc21-faastlane}, sharing the same concern, employs Intel Memory Protection Keys to ensure secure execution within threads and reduce function interaction costs. } 

% \qf{Sonic~\cite{atc21-sonic} and Faastlane~\cite{atc21-faastlane} focus on the overhead caused by data interactions between functions, and optimize function scheduling accordingly.}

% \qf{Wukong~\cite{socc20-wukong} and FaasFlow~\cite{asplos22-faasflow} highlight the overhead caused by the centralized scheduler and employ distributed sub-schedulers on worker nodes.}

% \qf{ORION~\cite{osdi22-orion} system emphasizes the optimization of serverless DAG workflows. It leverages the performance distribution of serverless functions to optimize the resource allocation for each function during the workflow's execution.} 

% \qf{Llama~\cite{socc21-llama}  is primarily focused on video processing, gathering statistics for each function to provide the optimal configuration for the workflow.}

% 

% \input{section/09-discuss}
\section{Conclusion}
\label{sec:conclusion}
In this paper, we identified the resource inefficiency lying in the early-binding based resource allocation for serverless workflows, and proposed a late-binding approach to address it by promoting bilateral runtime resource adaptation engaging both the developer and the provider.
Based on this concept, we proposed \namex---a novel resource adaptation framework for serverless workflows. 
We identified the challenges in building \namex and proposed efficient algorithms for fine-grained resource allocation %with MPS%
for \namex.
Experiments based on a system prototype show that \namex achieves significant resource savings while providing latency SLO guarantee.
Future work includes adding support for more complex workflows and exploring the impact of the runtime resource adaptation on function caching strategies.

% In this paper, we identified the DNN fragment misalignment problem in inference serving for hybrid DL and proposed a new concept called re-alignment to address it by promoting request batching and sharing.
% We proposed \namex---a first-of-its-kind inference serving system for hybrid DL adopting re-alignment. 
% We identified the challenges in building \namex and proposed efficient algorithms for fine-grained resource allocation %with MPS%
% for \namex.
% Experiments based on a system prototype show that \namex achieves significant resource savings while providing latency SLO guarantee.

\section*{Acknowledgment}
This work was supported in part by National Science Foundation of China under grant 62232012, in part by National Key Research \&
Development (R\&D) Plan under grant 2022YFB4501703, in part by the Major Key Project of PCL under Grant PCL2024A06 and PCL2022A05, and in part by the Shenzhen Science and Technology Program under Grant RCJC20231211085918010.
% The preferred spelling of the word ``acknowledgment'' in America is without 
% an ``e'' after the ``g''. Avoid the stilted expression ``one of us (R. B. 
% G.) thanks $\ldots$''. Instead, try ``R. B. G. thanks$\ldots$''. Put sponsor 
% acknowledgments in the unnumbered footnote on the first page.

\bibliographystyle{IEEEtran}
\bibliography{ref}

\end{document}